\documentclass[prd,twocolumn,floats,floatfix,nofootinbib]{revtex4}
\usepackage[dvips]{graphicx}
\usepackage{graphics}
\usepackage{dcolumn}
\usepackage{amssymb}
\usepackage{bm}
\usepackage{amsmath}
\usepackage{color}
\usepackage{subfigure}
\usepackage{epstopdf}

\begin{document}

\title{Relativistic Einstein Rings of Reissner-Nordstr\"om Black Holes \\
Nonminimally Coupled to Electrodynamics }

\author{Rodrigo Maier\footnote{rodrigo.maier@uerj.br}} 

\affiliation{
Departamento de F\'isica Te\'orica, Instituto de F\'isica, Universidade do Estado do Rio de Janeiro,\\
Rua S\~ao Francisco Xavier 524, Maracan\~a,\\
CEP20550-900, Rio de Janeiro, Brazil
}

\date{\today}

\begin{abstract}
In this paper we examine the relativistic Einstein rings
assuming a nonminimal coupling between gravitation and electromagnetism in a Reissner-Norstr\"om background.
Starting from a general action of a nonminimal coupled electrodynamics
we show that an unstable effective photon sphere may be obtained in the regime of eikonal approximation. 
Restricting ourselves to the unstable photon sphere domain we examine the expected angular positions of the first and second relativistic Einstein rings.
To compare our results with previous studies in the literature we model the
lens as a Galactic supermassive black hole. For fixed coupling parameters we show that such angular positions decrease as the charge parameter increases.
The angular separation between the first and second rings is also evaluated. 
We show that such separation increases as the charge parameter increases.
These patterns are not followed by nearly extremal configurations. In this case we show that there is an overlap domain so that the angular position and the corresponding coupling parameter do not allow one to differ extremal cases from complementary configurations which satisfy the cosmic censorship hypothesis.
\end{abstract}
\maketitle
\section{Introduction}
\label{intro}

Since the advent of black hole classical solutions\cite{Schwarzschild:1916uq,Reissner:1916cle,nord,Kerr:1963ud,Newman:1965my}
the behaviour of photon dynamics in a high energy regime has been exhaustively studied.
In this context, the deflection of light rays -- due to strong gravitational field -- with impact factor of the order of the black hole photon sphere may furnish attractive 
configurations in which high energy physics can be put to the test.
In this case, the position of the source with respect to the optical axis plays a significant role by defining the lensing configuration. In the minimal coupling case several studies addressed the lensing at large deflection angles\cite{Perlick:2004tq,Bozza:2010xqn,Virbhadra:1999nm,Virbhadra:2008ws,Eiroa:2002mk,Hsieh:2021scb}.
In this scenario, photons may reach a region sufficiently close to the unstable photon sphere so that they may go around the lens a number of times. 
For the case in which the source, lens and observer are aligned, an infinite
number of relativistic Einstein rings may be formed due to
the bending of light rays larger than $2\pi$. In the misaligned case on the other hand, an infinite sequence of relativistic images is produced on
both sides of the optical axis. 

Extending the above mentioned systems, the lensing at large deflection angles
has also been examined\cite{Bergliaffa:2020ivp} in the context of a nonminimal coupling between gravitation and the electrogmanetic field. From a theoretical ground there are several reasons behind the assumption of such nonminimal coupling. Among them we can mention the obtention of asymptotically flat black hole solutions with a positive ADM mass\cite{Mueller-Hoissen:1988cpx}, the effect of vacuum polarization on magnetic fields around a static black hole\cite{Pavlovic:2018idi} and exact 
cosmological solutions which describe isotropization processes\cite{Balakin:2005xi}.
For a complete review on general couplings between the electromagnetic
field and gravitation see \cite{Balakin:2005fu} and references therein. In this paper
we intend to extend the results obtained in \cite{Bergliaffa:2020ivp} considering the case in which the lens is described by a Galactic supermassive black hole with a nonvanishing charge parameter.
 
Although it is believed that charged black holes are rather unlikely to be observed in nature once they tend to attract opposite charges from their neighbourhood to neutralize themselves, a number of works have been developed in order to better understand 
proper mechanisms of charged black hole neutralization (see e.g.
\cite{Eardley:1975kp,Ruffini:2009hg,Hwang:2010im,Gong:2019aqa} and references therein).
From an astrophysical point of view it has been argued that 
binary black holes may admit electric charge. In fact, according to Zhang's mechanism\cite{Zhang:2016rli} a rotating
charged black hole may develop a magnetosphere which should allow a nonvanishing charge
parameter for a large period of time. The overall effect of such mechanism could provide a signature of an electromagnetic signal in gravitational wave events of binary black hole merger. Further analyses have given support to electric charges in black holes considering electromagnetic counterparts of black hole mergers\cite{Liebling:2016orx,Liu:2016olx,Punsly:2016abn,Levin:2018mzg,Deng:2018wmy}.
In the context of this paper we aim to examine the simplest configuration 
in which a nonrotating black hole with nonvanishing charge parameter changes the angular position of relativistic Einstein rings considering a nonminimal coupling between gravitation and the electromagnetic field.

We organize the paper as follows. In the next section we obtain an effective metric
from a Reissner-Nordst\"om background nonminimal coupled to electrodynamics in the regime of eikonal approximation. In Section III effective photon spheres are obtained.
In Section IV we obtain the angular position of the first and second Einstein relativistic rings
modeling our lens by a Galactic supermassive black hole with a charge parameter. In Section V we leave our final remarks.

\section{Effective Geometry from Nonminimal Coupling}
\label{sec:1}

We start by considering an action of a general nonminimal 
coupled electrodynamics
\begin{eqnarray}
\nonumber
S_\gamma&=&\int \sqrt{-g} F_{\mu\nu}F^{\mu\nu} d^4x   \\
&&+ \int \sqrt{-g} (\gamma_1 R F_{\mu\nu}F^{\mu\nu}+\gamma_2 R_{\mu\nu}F^{\mu}_{~~\beta}F^{\nu\beta}\\
\nonumber
&&~~~~~~~~~~~~~~~~~~~~~~+\gamma_3 R_{\mu\nu\beta\sigma}F^{\mu\nu}F^{\beta\sigma})d^4x,
\end{eqnarray}
where $R_{\mu\nu\beta\sigma}$ is the Riemann tensor -- 
$R_{\mu\nu}$ and $R$ are the Ricci tensor and the Ricci scalar, respectively -- $\gamma_i$ ($i=1, 2, 3$) are coupling coefficients and 
$F_{\mu\nu}=A_{\nu,\mu}-A_{\mu,\nu}$ is the Faraday tensor with $A_\mu$ as the potential $4$-vector.

Variations of $S_\gamma$ with respect $A_\mu$ yields the following equations of motion
\begin{eqnarray}
\label{eq2}
\nonumber
\nabla_{\mu} F^{\mu\nu}+\nabla_\mu \Big[\gamma_1 R F^{\mu\nu}+\frac{\gamma_2}{2}(R^\mu_{~~\beta}F^{\beta\nu}-R^\nu_{~~\beta}F^{\beta\mu})\\
+\gamma_3 R^{\mu\nu}_{~~~\beta\sigma}F^{\beta\sigma}\Big]=0,
\end{eqnarray}
where $\nabla_\alpha$ denotes the covariant derivative.
Finally, from Bianchi identities
we obtain the auxiliary conditions
\begin{eqnarray}
\label{eq3}
\nabla_{\alpha}F_{\mu\nu}+\nabla_{\mu}F_{\nu\alpha}+\nabla_{\nu}F_{\alpha\mu}=0.
\end{eqnarray}

In the following we aim to examine the photon dynamics which emerge from
(\ref{eq2})--(\ref{eq3}) in a high energy domain engendered from a black hole background.
In this context, 
the propagation of small perturbations of the electromagnetic field
around such background can be studied by means of the eikonal approximation 
%
in which the test electromagnetic field is given by
\begin{eqnarray}
\label{eq4}
F_{\mu\nu}=f_{\mu\nu}e^{i\theta}.
\end{eqnarray}
In the above, $\theta$ stands for a very rapidly varying phase
compared to the amplitude 
$f_{\mu\nu}$ when one takes into account scales much 
higher than the Compton wavelength of the electron. 
Defining $k_\mu=\theta_{,\mu}$, equation (\ref{eq2}) furnishes
\begin{eqnarray}
\label{eq5}
\nonumber
k_\mu (1+\gamma_1 R)f^{\mu\nu}+\frac{\gamma_2}{2}k_\mu(R^\mu_{~\beta}f^{\beta\nu}-R^\nu_{~\beta}f^{\beta\mu})\\
~~~~~~~~~~~~~~~~+\gamma_3 k_\mu R^{\mu\nu}_{~~~\beta\sigma}f^{\beta\sigma}=0.
\end{eqnarray}  
On the other hand, from (\ref{eq3}) we obtain
\begin{eqnarray}
\label{eq6}
k_{\alpha}f_{\mu\nu}+k_{\mu}f_{\nu\alpha}+k_{\nu}f_{\alpha\mu}=0,
\end{eqnarray}
so that
\begin{eqnarray}
k^2 f_{\mu\nu}+k_\alpha(k_\nu f^{\alpha}_{~\mu}-k_\mu f^{\alpha}_{~\nu})=0,    
\end{eqnarray}
where $k^2\equiv k^\alpha k_\alpha$.

Substituting (\ref{eq5}) in the above, we end up with
\begin{eqnarray}
\label{eq7}
\nonumber
k^2f_{\mu\nu}+\frac{k_\alpha}{1+\gamma_1 R}\Big\{
\frac{\gamma_2}{2} [(k_\nu R_{\mu\beta}-k_\mu R_{\nu\beta})f^{\beta\alpha}\\
+R^\alpha_{~\beta}(k_\mu f^{\beta}_{~\nu}-k_\nu f^{\beta}_{~\mu})]\\
\nonumber
+\gamma_3 (k_\mu R^\alpha_{~\nu\beta\sigma}-k_\nu R^\alpha_{~\mu\beta\sigma})f^{\beta\sigma}\Big\} =0.
\end{eqnarray}

Using standard coordinates $x^\alpha=(t, r, \theta, \phi)$
the Reissner-Nordstr\"om line element can be written 
in the base of 1-forms as
\begin{eqnarray}
ds^2=\eta_{AB}\Theta^A \Theta^B,    
\end{eqnarray}
where
\begin{eqnarray}
\Theta^A=e^A_{~\alpha}dx^\alpha,    
\end{eqnarray}
and
\begin{eqnarray}
e^A_{~\alpha}\rightarrow{\rm diag}\Big(f(r), \frac{1}{f(r)}, r, r\sin\theta\Big),
\end{eqnarray}
with
\begin{eqnarray}
f(r)\equiv \sqrt{1-\frac{2M}{r}+\frac{Q^2}{r^2}}.    
\end{eqnarray}
It is then easy to show that the 
Riemann tensor reads
\begin{eqnarray}
\nonumber
R^\alpha_{~\beta\gamma\delta}=h_1(r)\{\delta^\alpha_\gamma g_{\beta\delta}-\delta^\alpha_\delta g_{\beta\gamma}~~~~~~~~~~~~~~~~\\
\label{riemann}
+3[h_2(r)V^\alpha_{~\beta}V_{\gamma\delta}-h_3(r)W^\alpha_{~\beta}W_{\gamma\delta}]\},    
\end{eqnarray}
where
\begin{eqnarray}
&&V^{AB}=\sqrt{-g^{tt}g^{rr}}(e^A_{~t}e^B_{~r}-e^A_{~r}e^B_{~t}),\\
&&W^{AB}=\sqrt{g^{\theta\theta}g^{\phi\phi}}(e^A_{~\theta}e^B_{~\phi}-e^A_{~\phi}e^B_{~\theta}),
\end{eqnarray}
and 
\begin{eqnarray}
&&h_1(r)=\frac{M}{r^3}-\frac{Q^2}{r^4},\\
&&h_2(r)=1+\frac{Q^2}{3(Q^2-Mr)},\\
&&h_3(r)=1-\frac{Q^2}{3(Q^2-Mr)}.
\end{eqnarray}
The substitution of (\ref{riemann}) in (\ref{eq7}) -- and further contraction with $V^{\mu\nu}$ -- furnishes
a modified light cone\cite{Drummond:1979pp,Bergliaffa:2020ivp} 
dictaded by
\begin{eqnarray}
(1-\Sigma)(k_tk^t+k_rk^2)+k_\theta k^  \theta +k_\phi k^\phi=0.    
\end{eqnarray}
In this case, photons are expected to follow an effective geometry\cite{Barcelo:2005fc} 
whose line element reads
%
%
%
%
%
\begin{eqnarray}
\nonumber
\label{efg}
d\tilde{s}^2&=&\tilde{g}_{\mu\nu}dx^\mu dx^\nu\\
\label{em}
&=&\Big(1-\frac{1}{x}+\frac{q^2}{x^2}\Big)(1-\Sigma)dT^2\\
\nonumber
&&-\Big(1-\frac{1}{x}+\frac{q^2}{x^2}\Big)^{-1}(1-\Sigma)dx^2\\
\nonumber
&&~~-x^2(d\theta^2+\sin^2\theta d\phi^2),    
\end{eqnarray}
where
\begin{eqnarray}
T=\frac{t}{2M},~~x=\frac{r}{2M},~~q=\frac{Q}{2M},\\
\Gamma_2=\frac{\gamma_2}{(2M)^2},~~\Gamma_3=\frac{\gamma_3}{(2M)^2},
\end{eqnarray}
and
\begin{eqnarray}
\label{sig}
\Sigma\equiv \frac{3x\Gamma_3-q^2(\Gamma_2+8\Gamma_3)}{x^4+\Gamma_3(x-2q^2)}.   
\end{eqnarray}

\section{Effective Photon Spheres}
\label{sec:1}

To proceed we now investigate the motion of photons subjected to the effective geometry (\ref{em}).
The equations of motion are
\begin{eqnarray}
\label{geo}
\frac{d^2x^\mu}{d\sigma^2}
+\tilde{\Gamma}^\mu_{~\alpha\beta}
\frac{dx^\alpha}{d\sigma}\frac{dx^\beta}{d\sigma}=0,   
\end{eqnarray}
where we regard $\tilde{\Gamma}^\mu_{~\alpha\beta}$ as the Christoffel connection built with the effective geometry $\tilde{g}_{\mu\nu}$ and $\sigma$ is a proper affine parameter.
To simplify our analysis we restrict ourselves orbits with initial conditions $\theta_0=\pi/2$ and
$d\theta/d\sigma|_0=0$.
In this case it can be shown that the dynamics is restricted in the equatorial plane and
one can obtain two constants of motion,
\begin{eqnarray}
E=\Big(1-\frac{1}{x}+\frac{q^2}{x^2}\Big)(1-\Sigma)\Big(\frac{dT}{d\sigma}\Big),~~L=x^2\Big(\frac{d\phi}{d\sigma}\Big),    
\end{eqnarray}
connected to the energy and angular momentum, respectively.

Substituting the above results in the first integral
\begin{eqnarray}
\label{lag}
\tilde{g}_{\mu\nu}\frac{dx^\mu}{d\sigma}\frac{dx^\nu}{d\sigma}=0,
\end{eqnarray}
we obtain
\begin{eqnarray}
(1-\Sigma)^2\Big(\frac{dx}{d\sigma}\Big)^2+U_{eff}(x)=E^2,   
\end{eqnarray}
where
\begin{eqnarray}
\label{ep}
U_{eff}(x)=\frac{L^2\Big(1-\frac{1}{x}+\frac{q^2}{x^2}\Big)(1-\Sigma)}{x^2}.    
\end{eqnarray}
The maximum of the effective potential $U_{eff}$ defines the photon sphere.
For $\Sigma=0$ we obtain 
\begin{eqnarray}
x_{RN}=\frac{3+\sqrt{9-32q^2}}{4}    
\end{eqnarray}
denoting the Reissner-Nordstr\"om unstable photon sphere as one should expect.
Moreover, defining 
\begin{eqnarray}
x_h=\frac{1+\sqrt{1-4q^2}}{2}    
\end{eqnarray}
as the Reissner-Nordstr\"om event horizon, it can be easily shown from (\ref{ep}) that
\begin{eqnarray}
U_{eff}(x_h)=0.   
\end{eqnarray}
Furthermore, once
\begin{eqnarray}
\lim_{x\rightarrow +\infty}U_{eff}(x)=0    
\end{eqnarray}
and $dU_{eff}/dx\simeq -2L^2/x^3 < 0$ for large $x$
one may notice that $U_{eff}$ has at least one unstable photon sphere analogous to that of the 
Reissner-Nordstr\"om spacetime. In fact, for a negligible charge and small coupling parameters $\Gamma_2$ and 
$\Gamma_3$ this is the sole formed photon sphere. 
To see this behaviour, let us consider
an expansion of $U_{eff}$ up to first order in the coupling parameters. That is,
\begin{eqnarray}
\nonumber
U_{eff}(x)\simeq \Big(\frac{L}{x^3}\Big)^2\Big(1-\frac{1}{x}+\frac{q^2}{x^2}\Big)~~~~~~~~~~~~~~\\
~~~~~~~~~~~~~~~~\times[x^4-3x\Gamma_3+q^2(\Gamma_2+8\Gamma_3)].    
\end{eqnarray}
Assuming a sufficiently small charge and coupling parameters, namely $q$, $\Gamma_2$ and $\Gamma_3\ll1$, it can be shown that
\begin{eqnarray}
\frac{dU_{eff}}{dx}\simeq -L^2\Big[\frac{2x^6-3x^5+4q^2x^4-3\Gamma_3x^2(5x-6)}{x^9}\Big]. 
\end{eqnarray}
\begin{figure}
\includegraphics[width=8cm,height=5cm]{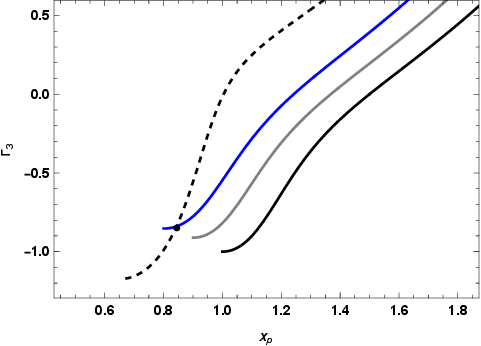}
\caption{$\Gamma_3$ as a function of $x_p$ for different charge parameters in the domain
$x>x_h$. The solid black, gray and blue curves are connected to $q=0$, $q=0.3$ and $q=0.4$, respectively. The black dashed curve on the other hand is connected to a nearly extremal configuration in which $q\simeq 0.5$. In this case one may notice
a degenerate behaviour. That is, 
for $\Gamma_3\simeq -0.85$, an unstable photon sphere located at $x_p\simeq 0.84$
(the black dot above)
can be obtained for different charge parameters, namely $q\simeq 0.5$ and $q=0.4$. 
}
\label{fig3}
\end{figure}
In this approximation it is then easy to show that the effective potential
has one global maximum connected to the effective unstable photon sphere located at
\begin{eqnarray}
\tilde{x}_p={x}_{RN}+\delta    
\end{eqnarray}
where 
\begin{eqnarray}
\delta=\frac{3x_{RN}(5x_{RN}-6)\Gamma_3}{x_{RN}^3(4x_{RN}-3)+18(5x_{RN}-7)\Gamma_3}.    
\end{eqnarray}

A degenerate behaviour may be notice considering the case of nearly extremal configurations and smaller coupling parameters. 
In fact, fixing $\Gamma_2=0$ for instance, the equation
\begin{eqnarray}
\label{eq0609}
\frac{dU_{eff}}{dx}\Big|_{x_p}=0    
\end{eqnarray}
furnishes the coupling parameter $\Gamma_3$ as a function of the photon sphere position $x_p$ and the charge parameter $q$. In Fig. 1 we show the behaviour of $\Gamma_3$ as a function of $x_p$ according to (\ref{eq0609}) for different charge parameters.
The black, gray and blue curves -- configurations far from the extremal case -- furnish different domains in which an unstable photon sphere is formed as one should expect.
However, as one moves towards nearly extremal configurations -- as 
the black dashed curve with $q\simeq 0.5$ in Fig. 1 -- an overlap domain in may be detected. That is, fixing a coupling parameter $\Gamma_3$, for instance, the same photon sphere can be obtained for different charge parameters. We illustrate an example of this case by the black dot in Fig. 1.

\section{Relativistic Einstein Rings}

We now examine the strong gravitational lensing connected to the effective geometry (\ref{efg}).
To this end we assume that photons reach a region sufficiently
close to the outer unstable photon
sphere so that they may go around
the lens an number of times. Considering that the source,
lens and observer are aligned, an infinite number of relativistic
Einstein rings are formed due to the bending of light rays larger than $2\pi$. In order to provide a model to sketch the source, lens and observer configuration we are going to adhere to the lens equation proposed by K. S. Virbhadra and G. F. R. Ellis\cite{Virbhadra:1999nm}. In this context we shall assume that
both observer and source are located in asymptotically flat regions
sufficiently far from the lens. The line connecting the observer, the lens and the source defines the optical axis.
Denoting $\beta$ and $\theta$ as the angle of the source and its image with respect to the optic axis, and $\alpha$ as the Einstein deflection angle, the lens equations reads  
\begin{eqnarray}
\tan\beta=\tan\theta-\sigma    
\end{eqnarray}
where
\begin{eqnarray}
\sigma=\frac{D_{LS}}{D_S}[\tan\theta+\tan(\alpha-\theta)].    
\end{eqnarray}
In the above
$D_{LS}$ is the lens-source distance and $D_S$ is the distance of the source from the observer. In this context the impact factor $J$ can be written as\cite{Virbhadra:1999nm,Bergliaffa:2020ivp}
\begin{eqnarray}
J=D_L \sin\theta.    
\end{eqnarray}
On the other hand, according to standard calculations the Einstein deflection angle in our case is given by
\begin{widetext}
\begin{eqnarray}
\alpha=2\int_{x_c}^\infty\frac{\sqrt{1-\Sigma(x)}dx}{x\sqrt{\frac{x^2}{x_c^2}\Big(1-\frac{1}{x_c}+\frac{q^2}{x_c^2}\Big)\frac{1-\Sigma(x_c)}{1-\Sigma(x)}-\Big(1-\frac{1}{x}+\frac{q^2}{x^2}\Big)}}-\pi,    
\end{eqnarray}
\end{widetext}
while the impact factor reads
\begin{eqnarray}
J=\frac{2M x_c}{\sqrt{\Big(1-\frac{1}{x_c}+\frac{q^2}{x_c^2}\Big)(1-\Sigma(x_c))}}.    
\end{eqnarray}
In the above $x_c$ is the closest distance of approach.
For the aligned case $\beta=0$ so that the lens equation reduces to
\begin{eqnarray}
\tan\theta=\frac{D_{LS}}{D_S}[\tan\theta+\tan(\alpha-\theta)].    
\end{eqnarray}

\begin{figure}
\includegraphics[width=8cm,height=5cm]{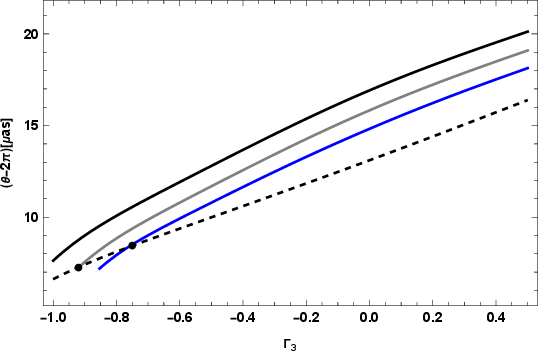}\\
\includegraphics[width=8cm,height=5cm]{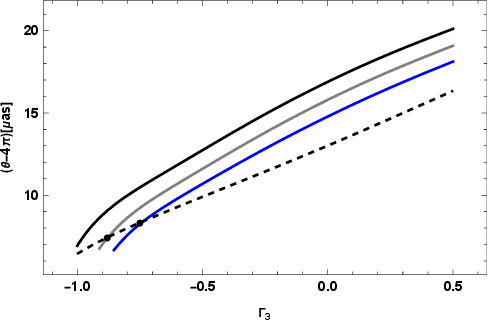}\\
\caption{The angular positions of the first (top panel) and second (bottom panel) 
relativistic Einstein rings as a function of $\Gamma_3$ for several charge parameters.
The solid black, gray and blue curves are connected to $q=0.0$, $q=0.3$ and $q=0.4$, respectively. Here we see that the angular position decreases as the charge parameter increases. The black dashed curve is connected to the nearly extremal configuration in which $q\simeq 0.5$. The black solid dots are examples of points of the domain of overlap of nearly extremal configurations and the complementary ones. }
\label{fig3}
\end{figure}

\begin{figure}
\includegraphics[width=8cm,height=5cm]{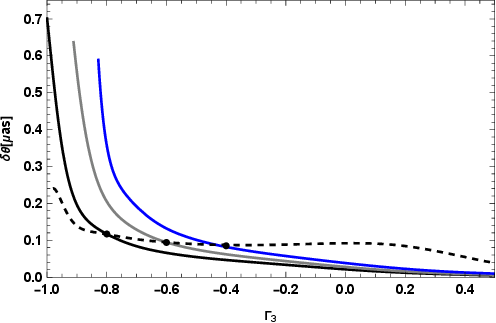}
\caption{The separation parameter $\delta\theta$ as a function of $\Gamma_3$ for several charge parameters. The solid black, gray and blue curves are connected to $q=0.0$, $q=0.3$ and $q=0.4$, respectively. Here we see that the separation parameter increases as the balck hole charge increases. The black dashed curve is connected to the nearly extremal configuration in which $q\simeq 0.5$. Again, the black solid dots are examples of points of the domain of overlap of nearly extremal configurations and the complementary ones.}
\label{fig3}
\end{figure}

To compare our results with those obtained in the literature\cite{Virbhadra:1999nm,Bergliaffa:2020ivp}
we shall model the lens as a Galactic supermassive black hole
with a nonvanishing charge. 
To this end we shall fix the parameters
\begin{eqnarray}
D_L=8.5{\rm Kpc},~~M=2.8\times 10^6 M\textsubscript{\(\odot\)},~~D_S=2D_{LS}    
\end{eqnarray}
where $M\textsubscript{\(\odot\)}$ is the Solar mass. Moreover, in order to simplify our analysis from now on we shall fix $\Gamma_2=0$. In fact, according to (\ref{sig}) one may notice that the coupling parameter $\Gamma_2$
plays a role of an additive constant -- together with $\Gamma_3$ -- connected to the black hole charge. Bearing this feature in mind we expect that the following results should hold, from a qualitative point of view, for a nonvanishing $\Gamma_2$ and in a similar domain of $\Gamma_3$. 

Following the standard route shown in \cite{Virbhadra:1999nm,Bergliaffa:2020ivp} we are then in a position to evaluate the angular position of the first 
and second relativistic Einstein rings. In Fig. 2 
we show the angular positions of the first (top panel) and second (bottom panel) relativistic Einstein rings considering several charge parameters as a function of $\Gamma_3$. Here we see that the angular positions decrease as the charge parameter increases. 

Defining 
\begin{eqnarray}
\theta_{1}=\theta-2\pi~~{\rm and}~~\theta_2=\theta-4\pi    
\end{eqnarray}
as the angular position of the first and second relativistic Einstein ring respectively, we introduce the separation parameter as
\begin{eqnarray}
\delta\theta= \theta_{1}-\theta_{2}.    
\end{eqnarray}
In Fig. 3 we show that the separation parameter increases as the charge parameter increases. 

Based on the numerical simulations of Figs. 2 and 3 we see that the above mentioned patterns cannot be extended up to nearly extremal configurations. In fact, according to our numerical examples one may notice that there is a domain of overlap
between nearly extremal configurations and the complementary ones which respect the 
cosmic censorship hypothesis. Examples of such overlaps are illustrated in Figs. 2 an 3 by the solid black dots. In this case the angular position together with its correponding coupling parameter do not allow one to differ nearly extremal cases from configurations far from those. 

\section{Final Remarks}
\label{sec:4}

In this paper we have examined strong lensing configurations
assuming a nonminimal coupling between gravitation and electromagnetism in a 
Reissner-Norstr\"om background. From an eikonal approximation we show that photons
follow an effective geometry which allows the formation of an effective unstable photon sphere. In order to simplify our analysis we 
restrict ourselves to the case in which photons reach a sufficiently close distance from
the outer unstable photon sphere so that relativistic Einstein rings can be observed once the source, lens and the observer are aligned. To compare our results to those in the literature we model the lens as a Galactic supermassive black hole with a nonvanishing electric charge. For fixed coupling parameters we show that the angular positions
of the first and second Einstein relativistic rings decrease as the charge parameter increases. On the contrary, for fixed coupling parameters the separations between the first and second Einstein rings increase as the charge parameter increases.
We show that these patterns cannot be extended up to nearly extremal configurations. In this case we show that there is an overlap domain in which the angular position together with its corresponding coupling parameter do not fix the black hole charge.

From an observational point of view it is worth to remark that the highest resolution telescope available today\cite{eht} has a resolution of the order of $19 \mu{\rm as}$.
Taking into account our results shown in Sec. IV we notice that the actual angular positions of the first and second relativistic Einstein rings in Fig. $2$ may put a better constrain 
on the coupling parameter $\Gamma_3$ rather than their separations in Fig. $3$. 
An additional difficulty to fix such constrain should be a proper method to infer the overall black hole charge. Nevertheless, the inceptive results presented here show that it may be feasible in the future to put better limits on the coupling parameters. Of course
the theoretical estimates presented in this paper may improve
if different/more general configurations are considered.
In this context, different ratios $D_{LS}/D_S$ and/or
misaligned configurations could provide a better screening for the allowed values for the coupling parameters.

Finally, the analysis presented in our paper should be extended to
more general cases such as Kerr black holes. Configurations in which the lens is modeled by a boosted black hole\cite{Aranha:2021zwf} could also furnish a more realistic system to be faced to observations. We shall examine these subjects in our further research.


\end{document}